\documentclass[twocolumn]{autart}    

\usepackage{picins} 
                                     
\usepackage{natbib}

\usepackage[dvips]{epsfig}    

\usepackage{amsmath,amssymb,amsfonts}
\usepackage{algorithmic}
\usepackage{textcomp}

\usepackage[dvipsnames]{xcolor}
\usepackage{dsfont}
\usepackage{subfigure}
\usepackage{float,caption}
\usepackage{bm}
\usepackage{mathtools}
\usepackage{physics}
\usepackage{theoremref}

\usepackage{tabularx}

\DeclareMathOperator*{\diag}{\rm Diag}

\newcommand{\qedsymbol}{\rule{2mm}{2mm}}

\begin{document}

\begin{frontmatter}

\title{Optimal Unbiased Linear Sensor Fusion over Multiple Lossy Channels with Collective Observability\thanksref{footnoteinfo}}

\thanks[footnoteinfo]{The work by Y. Wu and L. Shi is supported by a Hong Kong RGC General Research Fund 16204218. The work of Y. Li was supported by National Natural Science Foundation of China (61890924, 61991404), and Liao Ning Revitalization Talents Program (XLYC1907087). This paper was not presented at any conference.}

\thanks[cauthor]{Corresponding author.}

\author[Paestum]{Yuchi~Wu}\ead{ywubj@connect.ust.hk},
\author[B]{Kemi~Ding\thanksref{cauthor}}\ead{kemi.ding@ntu.edu.sg},
\author[C]{Yuzhe~Li}\ead{yuzheli@mail.neu.edu.cn}, and
\author[Paestum]{Ling~Shi}\ead{eesling@ust.hk}

\address[Paestum]{Department of Electronic and Computer Engineering, Hong Kong University of Science and Technology, Hong Kong SAR}
\address[B]{School of Electrical and Electronic Engineering, Nanyang Technological University, Singapore}
\address[C]{State Key Laboratory of Synthetical Automation for Process Industries, Northeastern University, Shenyang 110004, China}

\begin{keyword}
State estimation, Kalman filter, sensor fusion, lossy channel, collective observability.
\end{keyword}

\begin{abstract}
In this paper, we consider optimal linear sensor fusion for obtaining a remote state estimate of a linear process based on the sensor data transmitted over lossy channels. There is no local observability guarantee for any of the sensors. It is assumed that the state of the linear process is collectively observable. We transform the problem of finding the optimal linear sensor fusion coefficients as a convex optimization problem which can be efficiently solved.  Moreover, the closed-form expression is also derived for the optimal coefficients. Simulation results are presented to illustrate the performance of the developed algorithm.
\end{abstract}

\end{frontmatter}

\section{Introduction}
\label{sec:introduction}
Wireless sensor networks (WSNs) have been widely adopted in industrial processes, agricultural irrigation, smart grids, etc. Recently, due to the advances in wireless communication technologies, especially the advent of~5G era~(\cite{shafi20175g}), the spectrum of physical networks can be utilized more efficiently. Hence, more devices can be accommodated, which establishes the foundations for a large-scale Internet-of-Things~(IoT)~(\cite{mekki2019comparative}), ultra-dense networks~(\cite{wang2018power}), etc.

In order to reliably operate a large-scale dynamic system, it is necessary to obtain an accurate real-time state estimate at each time step. For this purpose, preliminary works have investigated management of the channel resources~(\cite{eisen2019learning}) as well as design of the data aggregation~(\cite{he2019consensus}) from various sensors. In this paper, we focus on the fusion problem of data from multiple sensors observing a physical process in a distributed manner.

The techniques in sensor fusion were developed for practical needs in state estimation or hypothesis testing problems. Some earlier works focused on estimating static variables based on fusion of raw measurements. For example, decentralized state estimation was investigated by~\cite{xiao2008linear} and~\cite{behbahani2012linear}, where a static unknown variable is estimated over networked channels, and the power allocation is optimized. Moreover, linear fusion has also been investigated in the decision fusion of hypothesis testing problems, e.g., spectrum sensing in cognitive radio (CR) network~(\cite{quan2009optimal}).

Later, sensor fusion has been extended to state estimation of dynamic systems. Channel activation over a graph topology as well as sensor selection problem for remote estimation has been considered by~\cite{yang2015deterministic}.

Aside from fusing the raw measurements, an alternative approach is to fuse the pre-processed data. The motivation is to save the communication resources. Linear compression of the sensor data before transmission is proposed by~\cite{zhang2003optimal}. Along this direction, smart sensors are employed to recursively extract information from the raw observations of dynamic systems.

The foundation of optimal linear fusion for smart sensors is laid by~\cite{sun2004multi}. Later works by~\cite{chen2014distributed,chen2016distributed} extend it to cases with channel failures or bandwidth constraints. To avoid high computational complexity, \cite{wu2019efficient} proposed an efficient algorithm for linear state fusion with scalar fusion coefficients. However, all these works assumed local observability of the entire process.

For example, when a large-scale dynamic system is monitored by a sensor network with each sensor observing certain sub-components of the system state, it is no longer feasible to estimate the system state locally. This collectively observable scenario has been addressed by~\cite{liu2017secure}, where the centralized Kalman filter can be recovered by a linear combination of the local estimations on the observable subspace of each sensor. The subspace decomposition methods for state estimation is also investigated by~\cite{yu2019event}. Meanwhile, \cite{he2018consistent} considered a distributed state estimation problem with graph topology, and covariance intersection~(CI) fusion strategy is adopted to bound the estimation error. A more recent result on distributed estimation for WSN is proposed by \cite{talebi2019distributed}, where the optimal performance of a centralized Kalman filter is recovered through average consensus algorithms on the local state estimates and the local covariance information from the population of sensors with collective observability.

In this work, we consider linear state fusion over lossy networks with collective observability. With Kalman decomposition, the observable subspace of each sensor is identified. Compared with~\cite{liu2017secure}, local estimates on the observable subspace of the sensors are transmitted to the remote state estimator through unreliable channels, and the optimal fusion coefficients are determined online. Several challenges emerge in our problem setup:
\begin{itemize}
\item[(1)] When \textbf{collective observability} is assumed, a global state estimate is infeasible at each sensor locally. As a result, it is necessary to propose a framework where smart sensors can \textbf{pre-process} their raw measurements effectively before transmitting over \textbf{unreliable channels}, and the remote state estimator is capable of \textbf{recovering} a state estimate of the entire system;
\item[(2)] Due to the lack of local observability, each sensor is merely capable of obtaining a ``\textbf{reduced-order}" information on the system state, which causes the \textbf{singularity} of the collective covariance matrix, hence results in~\cite{sun2004multi,chen2014distributed} are not applicable. Alternative approaches are required to compute the optimal fusion coefficients;
\item[(3)] For linear state fusion, it is desirable to obtain a \textbf{closed-from} solution of the \textbf{optimal fusion coefficients} in order to analyze the performance and the variation of optimal coefficients given different system parameters.
\end{itemize}

To cope with these difficulties, we have made the following contributions:
\begin{itemize}
\item[(1)] To perform sensor fusion with collective observability, we adopt \textbf{Kalman decomposition} and subspace projection to \textbf{extract informative data} from the observation of each sensor, which \textbf{generalize} previous works on linear fusion based on smart sensors by~\cite{sun2004multi,chen2014distributed,chen2016distributed} with local observability;
\item[(2)] For online implementation of the linear sensor fusion over lossy channels, we \textbf{transform} the original optimization problem for calculating the optimal coefficients for unbiased linear state fusion as a \textbf{linear programming} (\textbf{\thref{THM:optimal_coeff}}), which features a \textbf{complexity} of~$O(n^5 N^{2.5})$, where~$n$ is the dimension of the state of the dynamic process and~$N$ is the number of sensors. The complexity of our solution is comparable to~$O(n^3 N^3)$ in locally observable scenarios considered by~\cite{sun2004multi,chen2014distributed}. Besides, this approach avoids the computation of matrix inversions. The \textbf{stability} of the fusion estimation is shown in \textbf{\thref{prop:stable}};
\item[(3)] In order to obtain a \textbf{closed-form} expression for the optimal fusion coefficients, we draw an \textbf{analogy} with our problem and the minimum-variance unbiased estimate of an unknown parameter. According to \textbf{Gauss-Markov theorem} with singular covariance by~\cite{albert1973gauss}, the closed-form optimal fusion coefficient is obtained (\textbf{\thref{THM:closed_form}}).
\end{itemize}

The remainder of this paper is organized as follows. In Section~II, preliminary results on state estimation over a communication network is given, and the optimal unbiased linear state fusion with collective observability is posed as an optimization problem. Section~III transforms it as a linear programming~(LP), and shows the stability of the proposed fusion scheme. Section~IV gives the closed-form fusion coefficients. Numerical results are given in Section~V to illustrate the performance, and conclusions are drawn in Section~VI.

\subsection*{Notations:}
Denote the space of $n \times n$ positive semi-definite matrices as $\mathbb{S}_+^n$, and the space of $n \times m$ real matrix as $\mathbb{R}^{n \times m}$. For any singular~$M \in \mathbb{R}^{m \times n}$, its Moore-Penrose pseudo-inverse is denoted as~$M^{\dagger} \in \mathbb{R}^{n \times m}$. An identity matrix in $\mathbb{R}^{n \times n}$ is denoted as $I_{(n)}$. We denote a block diagonal matrix with diagonal elements~$D_1,D_2,\ldots,D_l$ as~$\bm{\diag} (D_1,D_2,\ldots,D_l)$. An indicator function is defined as $\delta_{kj} =1$ if $k=j$ and $\delta_{kj}=0$ otherwise. Given an operator $\mathcal{T}: X \to X$, the operation of applying it recursively is denoted as $\mathcal{T}^m=\underbrace{\mathcal{T} \circ \mathcal{T} \circ \cdots \circ \mathcal{T}}_{m}$.

	In order to clarify the meanings of different variables in this paper, we have summarized them in a table in Appendix~\ref{appendix:notations}.

\section{Preliminaries}
\begin{figure}[H]
	\begin{center}
		\includegraphics[width=0.5\textwidth]{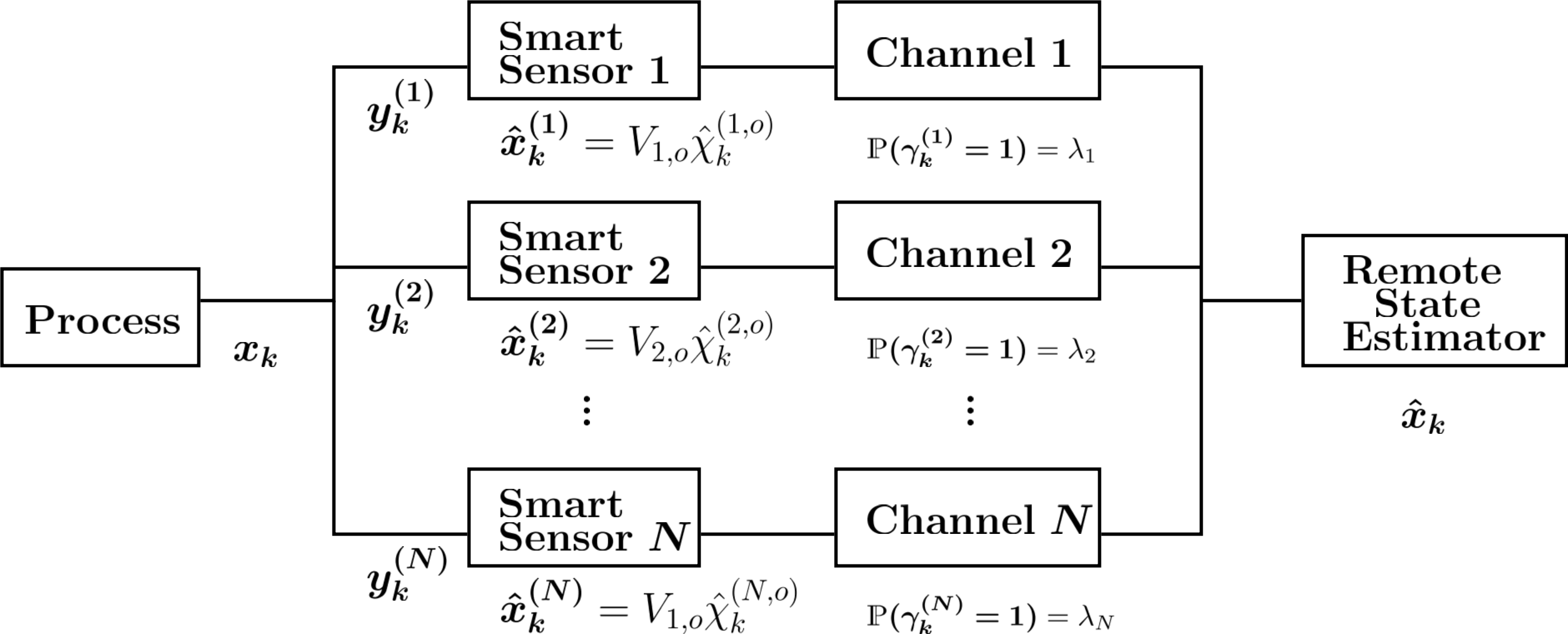}
		\caption{Remote state estimation of a linear process over multiple independent lossy channels.}
		\label{fig:system}
	\end{center}
\end{figure}

In this paper, we consider a group of~$N$ different smart sensors observing a linear process and forwarding their local state estimates to a remote state estimator through multiple independent lossy channels separately~(Fig.~\ref{fig:system}). The transmitted packets will be dropped with a certain probability, and the packet dropout for each channel follows an independent Bernoulli process.

At the other side of the channels, in order to obtain a global state estimate incorporating the received data, the remote state estimator employs a linear fusion scheme.

\subsection{Process and sensor models}
The process dynamic is given by
\begin{equation}
x_{k+1} = A x_k + w_k,
\label{eq:process}
\end{equation}
and each sensor takes its measurement independently with the following observation equation
\begin{equation}
y_k^{(i)} = C_i x_k + v_k^{(i)}.
\label{eq:observation}
\end{equation}
In~\eqref{eq:process} and~\eqref{eq:observation}, the state variable is $x_k \in \mathbb{R}^n$, and the observation of the state at time $k$ by sensor~$i$ is $y_k^{(i)} \in \mathbb{R}^{m_i}$. Assume the initial state variable $x_0 \in \mathbb{R}^n$ follows Gaussian distribution $\mathcal{N} ( \overline{x}_0, \Pi_0 )$ with $\Pi_0 \geq 0$. The Gaussian white noise in the system dynamics is $w_k \sim \mathcal{N}(0,Q)$ with $Q \geq 0$ and $\mathbb{E}[w_k w'_j] = 0$ for any $k \neq j$. The observation noise is $v_k^{(i)} \sim \mathcal{N} (0, R_i)$ with $R_i > 0$ and $\mathbb{E}[v_k^{(i)} v_j^{(i)'}] = 0$ for any $k \neq j$ and $i=1,2,\ldots,N$. It is assumed that $w_k$ and $v_k$ are uncorrelated. Besides, the initial state variable $x_0$ is independent of $w_k$. The observation noise process $v_k^{(i)}$ from different sensors are uncorrelated as well.


\begin{assum}
Assume~$Q>0$ and the system is collectively observable, i.e.,~$(A, \left[ C'_1 \ C'_2 \ \ldots \ C'_N \right]' )$ is observable.
\thlabel{assumption:ctrl_obser}
\end{assum}

\begin{rem}
Note that the assumption $Q > 0$ is slightly stronger than the conventional assumption of controllability of $(A, \sqrt{Q})$. This is to ensure that whenever the original state variable is projected onto the observable subspace of sensor $i$, which generally has a dimension smaller than $n$, controllability will still hold for the subsystem.
\thlabel{remark:controllability}
\end{rem}

For each sensor~$i$, it will first employ Kalman decomposition on $(A,C_i)$, and then the Kalman filter is adopted to obtain an optimal linear estimate of the state variable $x_k$ on the observable subspace of $(A, C_i)$. Denote the local state variable in the observable subspace of sensor~$i$ as ${\chi}_k^{(i)} \in \mathbb{R}^{n_i}$, where $n_i = \rank [C_i' \ \ A'C_i' \ \ A'^2 C_i' \ \ \ldots \ \ A'^{n-1} C_i']'$. According to Kalman decomposition, there exists an orthogonal coordinate transformation $T_i = [V_{i,\overline{o}} \ V_{i,o}]' \in \mathbb{R}^{n \times n}$ such that $T_i T_i' = I_n$. By defining ${\chi}_k^{(i)} := T_i x_k$ through a change of basis, an equivalent dynamic model of the linear process~\eqref{eq:process} is
\begin{equation}
{\chi}_{k+1}^{(i)} = T_i A T_i' {\chi}_{k}^{(i)} + T_i w_k =
\begin{bmatrix}
A_{i,\overline{o}} & \tilde{A}_i\\
O & A_{i,o}
\end{bmatrix} {\chi}_k^{(i)} + T_i w_k,
\label{eq:Kalman_decomposition}
\end{equation}
where $i = 1,2,\ldots,N$.

The locally observable state at sensor~$i$ can then be expressed as ${\chi}_k^{(i,o)} := V'_{i,o} x_k \in \mathbb{R}^{n_i}$ and its corresponding process noise is $\tilde{w}_k^{(i,o)} := V'_{i,o} w_k$. Thus, its dynamic model is
\begin{equation}
{\chi}_{k+1}^{(i,o)} = A_{i,o} {\chi}_k^{(i,o)} + \tilde{w}_k^{(i,o)}, \ i = 1,2, \ldots,N,
\label{eq:dynamic_observable}
\end{equation}
with the Gaussian white noise $\tilde{w}_k^{(i,o)} \sim \mathcal{N} (0,\tilde{Q}_{i,o})$, where $\tilde{Q}_{i,o} = V'_{i,o} Q V_{i,o}$. The initial state variable follows ${\chi}_0^{(i,o)} \sim \mathcal{N}(\overline{\chi}_0^{(i,o)}, \tilde{\Pi}_0^{(i,o)})$, where $\tilde{\Pi}_0^{(i,o)} \geq 0$.

The equivalent observation equation is
\begin{equation}
y_k^{(i)} = C_i x_k + v_k^{(i)} = C_i T_i' \cdot T_i x_k + v_k^{(i)}= \tilde{C}_i {\chi}_k^{(i,o)} + v_k^{(i)},
\end{equation}
where $\tilde{C}_i := C_i V_{i,o}$, and $i=1,2,\ldots,N$.

\begin{rem}
In the transformation matrix $T_i = [V_{i,\overline{o}} \ V_{i,o}]'$, the basis $V_{i,\overline{o}} \in \mathbb{R}^{n \times n_{i,\overline{o}}}$ consists of $n_{\overline{o}}$ column vectors forming an orthonormal basis of the unobservable subspace of $(A,C_i)$ while $V_{i,o} \in \mathbb{R}^{n \times n_{i,o}}$ consists of orthonormal basis vectors of the observable subspace, which are linearly independent of the basis of unobservable subspace. Thus, by definition of Kalman decomposition, the pair $(A_{i,o}, \tilde{C}_i)$ is observable for any sensor $i=1,2,\ldots,N$.
\end{rem}

In order for each sensor to locally obtain a state estimate, a Kalman filter is implemented for estimating~${\chi}_k^{(i,o)}$. For each sensor~$i$, it will treat its own observation $y_k^{(i)}$ of the linear process~\eqref{eq:process} as an input, and perform the following recursive update functions to obtain a local state estimate $\hat{\chi}_k^{s,(i)}$.
\begin{align}
\begin{cases}
\hat{\chi}_{k | k-1}^{s,(i)} &= A_{i,o} \hat{\chi}_{k-1}^{s,(i)}, \cr
P_{k|k-1}^{s,(i)} &= h_i (P_{k-1}^{s,(i)}), \cr
K_k^{(i)} &= P_{k | k-1}^{s,(i)} {\tilde{C}_i}' [\tilde{C}_i P_{k | k-1}^{s,(i)} {\tilde{C}_i}' + R_i]^{-1}, \cr
\hat{\chi}_k^{s,(i)} &= \hat{\chi}_{k | k-1}^{s,(i)} + K_k^{(i)} (y_k^{(i)} - \tilde{C}_i \hat{\chi}_{k|k-1}^{s,(i)}), \cr
P_k^{s,(i)} &= \tilde{g}_i (P_{k|k-1}^{s,(i)}),
\end{cases}
\label{eq:KF_update}
\end{align}
where $h_i$ and $\tilde{g}_i: \mathbb{S}_+^n \rightarrow \mathbb{S}_+^n$ are defined as follows:
\begin{align}
h_i (X) &\triangleq A_{i,o} X A_{i,o}' + \tilde{Q}_{i,o}, \label{func:h}\\
\tilde{g}_i (X) &\triangleq X-X \tilde{C}_i'[\tilde{C}_i X \tilde{C}_i'+R_i]^{-1} \tilde{C}_i X.
\end{align}

The recursion starts from $\hat{\chi}_0^{s,(i)}=\overline{\chi}_0^{(i)}$ and $P_0^{s,(i)} = \tilde{\Pi}_0^{(i)} \geq 0$, and the convergence result will be shown in \thref{lemma:converge_KF}.

\subsection{Communication over independent lossy channels}
Denote by~$\lambda_i \in (0,1]$ the packet arrival rate associated with sensor~$i$. For the packet containing $\hat{\chi}_k^{s,(i)}$ transmitted by sensor~$i$ at time~$k$, the arrival indicator is defined as
\begin{equation}
\gamma_k^{(i)} :=
\begin{cases}
1, & \hat{\chi}_k^{s,(i)} \text{arrives at the remote state estimator};\cr
0, & \text{Otherwise},
\end{cases}
\end{equation}
with~$\mathbb{P} \left( \gamma_k^{(i)} = 1 \right) = \lambda_i$.

\subsection{Sensor fusion at the remote state estimator}
Define the ``holding time" of each sensor as the number of consecutive time steps from the moment it receives its latest data packet to the current time~$k$, as expressed by
\begin{equation}
\tau_k^{(i)} := k - t_k^{(i)},
\end{equation}
where $t_k^{(i)} := \max \{t \leq k: \gamma_t^{(i)} = 1\}$.

Before we go into the details of the fusion methods, it is necessary to calculate the individual state estimates at the remote state estimator based on the received information set~$\mathcal{I}_k^{(i)}$. Denote the remote state estimate in the observable subspace of sensor~$i$ as~$\hat{\chi}_k^{(i,o)}$, then
\begin{align}
& \ \ \ \ \hat{\chi}_k^{(i,o)} := \mathbb{E} [\chi_k^{(i,o)} | \mathcal{I}_k^{(i)} ] = \mathbb{E} [\chi_k^{(i,o)} | \{ \hat{\chi}_{t_{\ell}^{(i)}}^{s,(i)}\}_{\ell=0}^k ]\nonumber\\
& = \mathbb{E} \bigg [A_{i,o}^{\tau_k^{(i)}} \chi^{(i,o)}_{k - \tau_k^{(i)}} + \sum\limits_{t=0}^{\tau_k^{(i)} - 1} A_{i,o}^{t} \tilde{w}^{(i,o)}_{k - t - 1} \bigg | \hat{\chi}_{k - \tau_k^{(i)}}^{s,(i)} \bigg ]\nonumber\\
& = A_{i,o}^{\tau_k^{(i)}} \hat{\chi}^{s,(i)}_{k - \tau_k^{(i)}}.
\label{eq:remote_individual_estimate}
\end{align}

Next, we project these individual estimates~$\hat{\chi}_k^{(i,o)}$ back into the original space of state variables~$x_k$. In particular, we fill the unobservable modes with zeros, then the projection is given by~$\hat{x}_k^{(i)} = T'_i \cdot [O \ \hat{\chi}_k^{(i,o)'}]' = V_{i,o} \hat{\chi}_k^{(i,o)} \in \mathbb{R}^n$.

We adopt a linear fusion scheme such that the local state estimates from different sensors are incorporated into an unbiased estimate of the state~$x_k \in \mathbb{R}^n$ at the remote state estimator, as follows.
\begin{equation}
\hat{x}_k= \sum\limits_{i=1}^N W_k^{(i)} \hat{x}_k^{(i)} = \sum\limits_{i=1}^N W_k^{(i)} V_{i,o} \hat{\chi}_k^{(i,o)},
\label{eq:linear_fusion}
\end{equation}
where we choose the coefficient matrix $W_k^{(i)}$ at each time $k \in \mathbb{N}_+$ to ensure that $\sum\limits_{i=1}^N W_k^{(i)} V_{i,o} V'_{i,o} = I_{(n)}$. This constraint gives an unbiased linear fusion. Moreover, we assume the packet dropout rate and the process dynamics satisfies the following assumption.
\begin{assum}
\begin{equation*}
\left(1 - \min\limits_{1 \leq i \leq N} \lambda_i \right) \rho^2 (A) < 1.
\end{equation*}
\thlabel{assumption:feasibility}
\end{assum}

\thref{assumption:feasibility} originates from~\cite{sinopoli2004kalman}, which ensures the stability of remote state estimation over a lossy channel.

\subsection{Error covariance of the fused state estimate}
Denote~$P_k$ as the estimation error covariance of~$\hat{x}_k$, which is defined by
\begin{equation}
P_k = \mathbb{E}[e_k e'_k],
\end{equation}
where~$e_k := x_k - \hat{x}_k$ is the error of the remote state estimate.

It is desirable to obtain a closed-form expression of~$P_k$ based on the linear fusion scheme~\eqref{eq:linear_fusion}. We firstly establish the convergence of the local Kalman filter by verifying the controllability of~$(A_{i,o}, \sqrt{\tilde{Q}_{i,o}})$ and the observability of~$(A_{i,o}, \tilde{C}_i)$~(\cite{kailath2000linear}).

\begin{lem}
Based on~\thref{assumption:ctrl_obser}, the pair $(A_{i,o}, \sqrt{\tilde{Q}_{i,o}})$ is controllable and~$(A_{i,o}, \tilde{C}_i)$ is observable for any~$i$. As a result, the estimation error covariance $P_k^{s,(i)}$ of the local Kalman filter~\eqref{eq:KF_update} converges exponentially to the steady-state error covariance~$\overline{P}^{(i,o)} > 0$.
\thlabel{lemma:converge_KF}
\end{lem}

\begin{pf}
We show the controllability of $\left( A_{i,o}, \sqrt{\tilde{Q}_{i,o}} \right)$ first. According to \thref{assumption:ctrl_obser}, $Q>0$. It remains to verify that~$\tilde{Q}_{i,o} := V_{i,o}' Q V_{i,o} \in \mathbb{R}^{n_{i,o} \times n_{i,o}}$ is positive definite.

Given the fact that $T_i T_i' = I$ for $T_i = \begin{bmatrix} V_{i,\overline{o}} & V_{i,o} \end{bmatrix}'$, it is straightforward to obtain that $V_{i,\overline{o}}' V_{i,\overline{o}} = I_{(n_{i,\overline{o}})}$ and $V_{i,o}' V_{i,o} = I_{(n_{i,o})}$, thus $\rank V_{i,\overline{o}} = n_{i,\overline{o}}$ and $\rank V_{i,o} = n_{i,o}$. Now we take an arbitrary but fixed non-zero vector $\bm{x}_{i,o} \in \mathbb{R}^{n_{i,o}} / \{\bm{0}\}$, and since $\norm{V_{i,o} \bm{x}_{i,o}}^2 = \bm{x}_{i,o}' V_{i,o}' V_{i,o} \bm{x}_{i,o} = \norm{\bm{x}_{i,o}}^2 >0$, a higher-dimensional non-zero vector can be obtained as $V_{i,o} \bm{x}_{i,o} \in \mathbb{R}^{n} / \{\bm{0}\}$. As a result, given $Q>0$, we can obtain by definition of a positive definite matrix that $\bm{x}_{i,o}' V_{i,o}' Q V_{i,o} \bm{x}_{i,o} > 0$ for any fixed $\bm{x}_{i,o} \in \mathbb{R}_{n_{i,o}} / \{\bm{0}\}$. Therefore, it is verified that $V_{i,o}' Q V_{i,o} > 0$.

Based on Popov-Belovich-Hautus (PBH) test, the controllability of $\left( A_{i,o}, \sqrt{\tilde{Q}_{i,o}} \right)$ is equivalent to 
\begin{equation*}
\rank \begin{bmatrix} A_{i,o} - \lambda I & \sqrt{\tilde{Q}_{i,o}} \end{bmatrix} = n_{i,o}, \ \forall \lambda \in \mathbb{R}.
\end{equation*}

We can then get~$n \geq \rank \begin{bmatrix} A_{i,o} - \lambda I & \sqrt{\tilde{Q}_{i,o}} \end{bmatrix} \geq \rank \sqrt{\tilde{Q}_{i,o}} = \rank \tilde{Q}_{i,o} = n_{i,o}, \ \forall \lambda \in \mathbb{R}$,
which indicates that $\left( A_{i,o}, \sqrt{\tilde{Q}_{i,o}} \right)$ is controllable.

The pair $\left( A_{i,o}, \tilde{C}_i \right)$ is observable due to the properties of Kalman decomposition. Hence, the controllability of $\left(A_{i,o}, \sqrt{\tilde{Q}_{i,o}}\right)$ as well as the observability of $\left( A_{i,o}, \tilde{C}_i \right)$ has been verified successfully.

Next, it comes to the convergence property of the local Kalman filters~\eqref{eq:KF_update} given the above results. From~\cite{kailath2000linear}, the controllability and observability shown above ensure the convergence of the local Kalman filter. Specifically, there is a unique fixed-point ${P^*}^{(i,o)} > 0$ for the Riccati equation $X = h_i \circ \tilde{g}_i (X)$, which corresponds to the steady-state prediction error. Then, the steady-state estimation error is $\overline{P}^{(i,o)} = \tilde{g}_i ({P^*}^{(i,o)})$.

As $R_i > 0$ for any~$i$, based on the information form of a Kalman filter and matrix inversion lemma, it can be obtained that
\begin{align*}
\overline{P}^{(i,o)} = \tilde{g}_i ({P^*}^{(i,o)})= \bigg [I + {P^*}^{(i,o)} \tilde{C}_i R_i^{-1} \tilde{C}_i \bigg ]^{-1} \cdot {P^*}^{(i,o)}.
\end{align*}

As the steady-state prediction error ${P^*}^{(i,o)} > 0$, we can obtain that $\overline{P}^{(i,o)} > 0$. \qedsymbol
\end{pf}

From~\thref{lemma:converge_KF}, we conclude that the optimal Kalman gain $K_k^{(i)}$ also converges to a constant value $K_i^* = {P^*}^{(i,o)} \tilde{C}'_i [\tilde{C}_i {P^*}^{(i,o)} \tilde{C}'_i + R_i]^{-1}$. For simplicity of analysis, we assume the local Kalman filters at the sensors have been operating for an adequately long time such that each has converged to its steady-state at~$k=0$.

As overlaps may exist among the observable subspaces of different sensors, the local state estimates obtained by each sensor may correlate with each other. Denote the error of local state estimator $i$ at time step $k$ is $\mathcal{E}_k^{s,(i)}={\chi}_k^{(i,o)}-\hat{\chi}_k^{s,(i)} \in \mathbb{R}^{n_{i,o}}$. Thus, it is also necessary to analyze the convergence of the cross-covariances between the estimation errors~$\mathcal{E}_k^{s,(i)}$ and~$\mathcal{E}_k^{s,(j)}$ with~$i \neq j$.

\begin{lem}
For each pair of different local state estimators $i, j \in \{1,2,\ldots,N\} \ (i \neq j)$, the recursive update of cross-correlation matrix $\Gamma_k^{ij}:=\mathbb{E}[\mathcal{E}_k^{s,(i)} \mathcal{E}_k^{s,(j)'}] \in \mathbb{R}^{n_{i,o} \times n_{j,o}}$ is
\begin{equation}
\Gamma_{k+1}^{ij} = \mathcal{T}_{ij} (\Gamma_k^{ij}),
\label{eq:correlation_update}
\end{equation}
where the mapping~$\mathcal{T}_{ij}: \mathbb{R}^{n_{i,o} \times n_{j,o}} \to \mathbb{R}^{n_{i,o} \times n_{j,o}}$ is given by~$\mathcal{T}_{ij} (X) = (I - K_i^* \tilde{C}_i ) h_{ij}(X) (I - K_j^* \tilde{C}_j)'$, with $h_{ij}: \mathbb{R}^{n_{i,o} \times n_{j,o}} \to \mathbb{R}^{n_{i,o} \times n_{j,o}}$ defined as $h_{ij} (X) := A_{i,o} X A'_{j,o} + V'_{i,o} Q V_{j,o}$. As~$k \to \infty$, when the local Kalman filters converge to their steady states, the cross correlation matrix also converges to a fixed point of the mapping~$\mathcal{T}_{ij}$, denoted by~$\overline{\Gamma}_{ij}$, i.e., $\lim\limits_{k \to \infty} \Gamma_k^{ij} = \overline{\Gamma}_{ij}$.
\thlabel{lemma:fixed_point_cross_covariance}
\end{lem}

\begin{pf}
First, we show the update function \eqref{eq:correlation_update}. According to the system model \eqref{eq:process}, \eqref{eq:dynamic_observable} and the update functions for the Kalman filter \eqref{eq:KF_update}, the estimation error at the local state estimator $i$ can be expressed as
\begin{align*}
& \ \ \mathcal{E}_k^{s,(i)} = {\chi}_k^{(i,o)} - \hat{\chi}_k^{s,(i)}\\
&=(I - K_i^* \tilde{C}_i) A_{i,o} \mathcal{E}_{k-1}^{s,(i)} + (I- K_i^* \tilde{C}_i) \tilde{w}_{k-1}^{(i,o)} - K_i^* v_k^{(i)}.
\end{align*}

Then the cross-correlation at time step $k$ can be expressed as
\begin{align}
& \Gamma_k^{ij} =(I - K_i^* \tilde{C}_i ) h_{ij}(\Gamma_{k-1}^{ij}) (I - K_j^* \tilde{C}_j)' \nonumber\\
&=(I - K_i^* \tilde{C}_i ) (A_{i,o} \Gamma_{k-1}^{ij} A'_{j,o} + V'_{i,o} Q V_{j,o}) (I - K_j^* \tilde{C}_j)' \nonumber\\
&= \mathcal{T}_{ij} \left( \Gamma_{k-1}^{ij} \right).
\label{eq:error_update}
\end{align}

Now, we pick two different initial values~$\Gamma_0^{ij}$ and~$\tilde{\Gamma}_0^{ij}$ for the cross-covariance between sensor~$i$ and~$j$. It can be obtained that
\begin{align*}
& \ \ \ \ \norm{\Gamma_k^{ij} - \tilde{\Gamma}_k^{ij}} = \norm{\mathcal{T}_{ij} (\Gamma_{k-1}^{ij}) - \mathcal{T}_{ij} (\tilde{\Gamma}_{k-1}^{ij})}\\
&= \norm{(A_{i,o} - K_i^* \tilde{C}_i A_{i,o}) (\Gamma_{k-1}^{ij} - \tilde{\Gamma}_{k-1}^{ij}) (A_{j,o} - K_j^* \tilde{C}_j A_{j,o})'}\\
&= \cdots\\
&= \norm{(A_{i,o} - K_i^* \tilde{C}_i A_{i,o})^k (\Gamma_0^{ij} - \tilde{\Gamma}_0^{ij}) (A_{j,o} - K_j^* \tilde{C}_j A_{j,o})'^k}\\
&\leq \norm{(A_{i,o} - K_i^* \tilde{C}_i A_{i,o})^k} \cdot \norm{\Gamma_0^{ij} - \tilde{\Gamma}_0^{ij}} \cdot\\
& \ \ \ \ \norm{(A_{j,o} - K_j^* \tilde{C}_j A_{j,o})^k}.
\end{align*}

According to Corollary~5.6.14 in~\cite{horn1990matrix}, for a matrix norm~$\norm{\cdot}$ and an arbitrary square matrix~$X \in \mathds{R}^{n \times n}$, there is~$\rho(X) = \lim\limits_{k \to \infty} \norm{X^k}^{\frac{1}{k}}$. In other words, for any~$\epsilon > 0$, there exists a~$K_0 > 0$ such that~$\abs{\norm{X^k}^{\frac{1}{k}} - \rho(X)} < \epsilon$ for all~$k > K_0$, which is equivalent to that for any~$k > K_0$,
\begin{equation}
(\rho(X) - \epsilon)^ k < \norm{X^k} < (\rho(X) + \epsilon)^k.
\label{eq:spectrum_radius}
\end{equation}

Take $\epsilon := \min\{\frac{1- \rho(A_{i,o} - K_i^* \tilde{C}_i A_{i,o})}{2}, \frac{1- \rho(A_{j,o} - K_j^* \tilde{C}_j A_{j,o})}{2},\\ \frac{\rho(A_{i,o} - K_i^* \tilde{C}_i A_{i,o})}{2}, \frac{\rho(A_{j,o} - K_j^* \tilde{C}_j A_{j,o})}{2}\}$, there is
\begin{align*}
\norm{\Gamma_k^{ij} - \tilde{\Gamma}_k^{ij}} &\leq [\rho(A_{i,o} - K_i^* \tilde{C}_i A_{i,o}) + \epsilon]^k \cdot \norm{\Gamma_0^{ij} - \tilde{\Gamma}_0^{ij}} \cdot\\
& \ \ \ \ [\rho(A_{j,o} - K_j^* \tilde{C}_j A_{j,o}) + \epsilon]^k, \ \forall k > K_0.
\end{align*}

Since by Kalman decomposition that $(A_{i,o}, \tilde{C}_i)$ is observable for all~$i$, under the steady-state optimal Kalman gain $K_i^*$, we must have $A_{i,o} -K_i^* \tilde{C}_i A_{i,o}$ stable, i.e., $\rho(A_{i,o} -K_i^* \tilde{C}_i A_{i,o})< 1$. Based on the chosen~$\epsilon$, there is also~$\rho(A_{i,o} -K_i^* \tilde{C}_i A_{i,o}) + \epsilon < 1$ and~$\rho(A_{j,o} -K_j^* \tilde{C}_j A_{j,o}) + \epsilon < 1$. As a result, we have~$\lim\limits_{k \to \infty} \norm{\Gamma_k^{ij} - \tilde{\Gamma}_k^{ij}} = 0$ for any initial values~$\Gamma_0^{ij}, \ \tilde{\Gamma}_0^{ij} \in \mathds{R}^{n_{i,o} \times n_{j,o}}$.

Thus, we can conclude that the sequence~$\{\Gamma_k^{ij}\}_{k \geq 0}$ converges. Its limit~$\overline{\Gamma}_{ij} \in \mathds{R}^{n_{i,o} \times n_{j,o}}$ is the steady-state cross-covariance between sensor~$i$ and~$j$. \qedsymbol
\end{pf}

Denote the individual estimation error corresponding to each sensor~$i$ as $\mathcal{E}_k^{(i,o)}:=\chi_k^{(i,o)} - \hat{\chi}_k^{(i,o)}$, and the associated error covariance matrix is
\begin{equation}
P_k^{(ii)} = \mathbb{E}[\mathcal{E}_k^{(i,o)} \mathcal{E}_k^{(i,o)'}] = h_i^{\tau_k^{(i)}} (\overline{P}^{(i,o)}),
\label{eq:err_cov}
\end{equation}
which only depends on the ``holding time" $\tau_k^{(i)}$ of state estimator~$i=1,2,\ldots,N$.

The cross correlation matrix~$P_k^{(ij)} := \mathbb{E}[\mathcal{E}_k^{(i,o)} \mathcal{E}_k^{(j,o)'}]$ can be expressed as

\begin{align}
P_k^{(ij)} &= \mathbb{E}[\mathcal{E}_k^{(i,o)} \mathcal{E}_k^{(j,o)'}]\nonumber\\
	&=\mathbb{E}[(\chi_k^{(i,o)} - \hat{\chi}_k^{(i,o)})(\chi_k^{(j,o)} - \hat{\chi}_k^{(j,o)})']\nonumber\\
	&=\mathbb{E}[(A_{i,o}^{\tau_k^{(i)}} \chi_{k-\tau_k^{(i)}}^{(i,o)} + \sum\limits_{t=0}^{\tau_k^{(i)}-1} A_{i,o}^t \tilde{w}_{k-t-1}^{(i,o)} - A_{i,o}^{\tau_k^{(i)}} \hat{\chi}_{k-\tau_k^{(i)}}^{s,(i)})\nonumber\\
	& \ \ \ \ \ \ (A_{j,o}^{\tau_k^{(j)}} \chi_{k-\tau_k^{(j)}}^{(j,o)} + \sum\limits_{t=0}^{\tau_k^{(j)}-1} A_{j,o}^t \tilde{w}_{k-t-1}^{(j,o)} - A_{j,o}^{\tau_k^{(j)}} \hat{\chi}_{k-\tau_k^{(j)}}^{s,(j)})']\nonumber\\
	&=\mathbb{E}[(A_{i,o}^{\tau_k^{(i)}} \mathcal{E}_{k-\tau_k^{(i)}}^{s,(i)} + \sum\limits_{t=0}^{\tau_k^{(i)}-1} A_{i,o}^t \tilde{w}_{k-t-1}^{(i,o)}) \cdot \nonumber\\
	& \ \ \ \ \ \ (A_{j,o}^{\tau_k^{(j)}} \mathcal{E}_{k-\tau_k^{(j)}}^{s,(j)} + \sum\limits_{t=0}^{\tau_k^{(j)}-1} A_{j,o}^t \tilde{w}_{k-t-1}^{(j,o)})']\nonumber\\
&=A_{i,o}^{\tau_k^{(i)}} \overline{\Gamma}_{ij} A'^{\tau_k^{(j)}}_{j,o} + \sum\limits_{t=0}^{\min\{ \tau_k^{(i)}, \tau_k^{(j)} \} -1} A^t_{i,o} V'_{i,o} Q V_{j,o} A'^t_{j,o},
\label{eq:cross_correlation}
\end{align}
where the third equation is based on~\eqref{eq:dynamic_observable} and~\eqref{eq:remote_individual_estimate}, the fifth equation holds as the Kalman filter~\eqref{eq:KF_update} reaches steady state.

This expression of~$P_k^{(ij)}$ depends on the holding time $\tau_k^{(i)}$ and $\tau_k^{(j)}$, and it holds for any $i \neq j, \ i, j \in \{1,2,\ldots,N\}$.


Denote the stacked fusion-coefficient matrix by~$\bm{W}_k := [W_k^{(1)} \ \ W_k^{(2)} \ \ \ldots \ \ W_k^{(N)}]' \in \mathbb{R}^{nN \times n}$ and~$\bm{V}_o := [V_{1,o} V'_{1,o} \ \ V_{2,o} V'_{2,o} \ \ \ldots \ \ V_{N,o} V'_{N,o}]' \in \mathbb{R}^{nN \times n}$. Suppose the fusion coefficients are chosen appropriately such that $\sum\limits_{i=1}^N W_k^{(i)} V_{i,o} V'_{i,o} = \bm{W}_k' \bm{V}_o = I_{(n)}$, i.e., it provides an unbiased linear fusion. Then, the error covariance of the fusion estimation~$\hat{x}_k$ is thus
\begin{align}
P_k = \mathbb{E}[e_k e'_k] 
= \sum\limits_{i=1}^N \sum\limits_{j=1}^N W_k^{(i)} V_{i,o} P_k^{(ij)} V'_{j,o} W'^{(j)}_k =\bm{W}'_k \Sigma \bm{W}_k,
\label{eq:error_remote}
\end{align}
in which the covariance matrix~$\Sigma \in \mathbb{R}^{nN \times nN}$ is
\begin{align}
&\Sigma := \nonumber \\
&\begin{bmatrix}
    V_{1,o} P_k^{(11)} V'_{1,o} & V_{1,o} P_k^{(12)} V'_{2,o} & \dots  & V_{1,o} P_k^{(1N)} V'_{N,o}\\
    V_{2,o} P_k^{(21)} V'_{1,o} & V_{2,o} P_k^{(22)} V'_{2,o} & \dots  & V_{2,o} P_k^{(2N)} V'_{N,o}\\
    \vdots & \vdots  & \ddots & \vdots \\
    V_{N,o} P_k^{(N1)} V'_{1,o} & V_{N,o} P_k^{(N2)} V'_{2,o} & \dots  & V_{N,o} P_k^{(NN)} V'_{N,o}
\end{bmatrix}.
\label{def:Sigma}
\end{align}

\subsection{Problem formulation}
The problem of optimal linear fusion at the remote state estimator is stated as an optimization problem which minimizes the estimation error~\eqref{eq:error_remote} by designing the fusion coefficient.

\begin{prob}[Unbiased linear state fusion]
\begin{align*}
& \min\limits_{\bm{W}_k \in \mathbb{R}^{nN \times n}} \ \tr(P_k),\\
& \ s.t. \ \bm{W}'_k \bm{V}_o = I_{(n)},
\end{align*}
where $\tr(P_k) = \tr(\bm{W}'_k \Sigma \bm{W}_k)$ according to~\eqref{eq:error_remote}.
\thlabel{problem:sensor_fusion}
\end{prob}

\section{Main results}
We transform \thref{problem:sensor_fusion} as a linear programming~(LP), for which efficient algorithms exist.

\subsection{Optimal linear fusion coefficients}
First, we show the positive semi-definiteness of the matrix~$\Sigma \in \mathbb{R}^{nN \times nN}$.

\begin{lem}
The matrix $\Sigma \in \mathbb{R}^{nN \times nN}$ in~\eqref{def:Sigma} is symmetric and $\Sigma \geq 0$.
\thlabel{lemma:Sigma_psd}
\end{lem}

See Appendix~\ref{appendix:proof_Sigma_psd} for proof.

In order to efficiently calculate the optimal fusion coefficients and to show the stability of the optimal linear fusion estimation, it is helpful to reformulate \thref{problem:sensor_fusion}. In order to do so, we introduce a new auxiliary variable and relax the constraints. The relaxed version of the original sensor fusion problem is given as follows.

\begin{prob}[Relaxed optimization problem]
\begin{equation}
\min_{\bm{W}_k \in \mathbb{R}^{nN \times n}, \bm{X}_k \in \mathbb{R}^{nN \times nN}} \ \tr \left( \Sigma \bm{X}_k \right),
\label{eq:obj_SDP}
\end{equation}
s.t.~$\bm{W}_k' \bm{V}_o = I_{(n)}$ and $\bm{X}_k \succeq \bm{W}_k \bm{W}_k'$.
\thlabel{problem:SDP}
\end{prob}

It can be verified that \thref{problem:SDP}, as a relaxed problem, has no loss of optimality compared to the original \thref{problem:sensor_fusion} on linear fusion. This result is stated in the following lemma, of which the proof establishes the necessity for the optimal solution pair to take equality in the constraint~$\bm{X}_k \succeq \bm{W}_k \bm{W}_k'$.

\begin{lem}
The optimal solution to \thref{problem:SDP} coincides with the optimal solution to \thref{problem:sensor_fusion}.
\thlabel{lemma:equivalence_fusion_SDP}
\end{lem}

See Appendix~\ref{appendix:proof_equivalence_fusion_SDP} for proof.

With help from \thref{lemma:equivalence_fusion_SDP}, we are able to build the bridge for the equivalence between \thref{problem:sensor_fusion} and a linear programming.

\begin{prob}[Transformed linear programming]
\begin{equation}
\max_{\bm{W}_k \in \mathbb{R}^{nN \times n}, \Lambda_1 \in \mathbb{R}^{n \times n}} \ \frac{1}{2} \ \tr \left( \Lambda_1 \right),
\end{equation}
s.t.~$\bm{W}_k' \bm{V}_o = I_{(n)}$ and~$\ 2 \Sigma \bm{W}_k = \bm{V}_o \Lambda_1'$.
\thlabel{problem:dual}
\end{prob}

\begin{thm}
The \thref{problem:sensor_fusion} for finding the optimal fusion coefficients can be solved through the linear programming in \thref{problem:dual}.
\thlabel{THM:optimal_coeff}
\end{thm}

See Appendix~\ref{appendix:proof_optimal_coeff} for proof.

Based on the complexity of linear programming as analyzed by \cite{vaidya1989speeding}, \thref{problem:dual} can be solved in polynomial time~$O(n^5 N^{2.5})$. Hence, with \thref{THM:optimal_coeff}, efficient algorithms exist for calculating the optimal fusion coefficients.

\subsection{Stability of the remote state estimate}
The stability of this remote fusion estimation is given in the following proposition.

\begin{prop}
Denote as~$P_k^*$ the error covariance of the remote state estimate under the optimal fusion coefficients~$\bm{W}_k^*$ given by the solution to~\thref{problem:sensor_fusion} at time~$k$. Then,
\begin{equation}
\lim\limits_{k \to \infty} \tr \left( \mathbb{E} [P_k^*] \right) < \infty.
\end{equation}
\thlabel{prop:stable}
\end{prop}

See Appendix~\ref{appendix:proof_stable} for proof.

\subsection{Closed-form optimal fusion coefficients by Gauss-Markov theorem}

In \thref{problem:dual}, we have formulated the problem of finding the unbiased linear state fusion coefficients as a linear programming (LP), which can be solved efficiently. Meanwhile, we are interested in obtaining a closed-form expression of the optimal fusion coefficients.

The classical parameter estimation problem given a noisy linear observation is analyzed by~\cite{albert1973gauss}, where the covariance matrix of the observation noise is possibly singular. The explicit parameter estimation problem is formulated as follows.

Consider observations of the form
\begin{equation}
z=Hx+v,
\end{equation}
where~$H \in \mathbb{R}^{n \times p}$ is the observation matrix, and the vector~$x \in \mathbb{R}^p$ is a constant but unknown vector to be estimated. The zero-mean observation noise~$v \in \mathbb{R}^n$ has a singular covariance matrix~$V \in \mathbb{R}^{n \times n}$ with~$V \geq 0$.

Now, we plan to find a linear estimate of the parameter~$x$ based on the noisy observation~$z \in \mathbb{R}^n$, i.e., to find the appropriate gain~$K \in \mathbb{R}^{p \times n}$ such that the estimate~$\hat{x} \in \mathbb{R}^n$ is
\begin{equation}
\hat{x} = K z.
\end{equation}

We hope to find an unbiased estimate which minimizes the estimation error, hence this can be formulated as a constrained optimization problem.

The objective can be expressed as
\begin{align*}
\mathbb{E} [\norm{\hat{x} - x}^2] &=\mathbb{E} [\norm{Kz - x}^2]\\
&=\mathbb{E} [\norm{(KH - I) x + Kv}^2]\\
&=\mathbb{E} [\norm{(KH - I) x}^2] + \tr(KVK').
\end{align*}

In order for the state estimate to be unbiased, it is necessary to have~$\mathbb{E}[\hat{x} - x] = 0$, i.e.,~$KH=I$, and the objective function becomes~$\tr \left( K V K' \right)$. According to~\cite{luenberger1997optimization}, the constrained optimization problem is given as follows.

\begin{prob}[Minimum-variance unbiased estimate]
\begin{align*}
& \min\limits_{K \in \mathbb{R}^{p \times n}} \tr \left( K V K' \right),\\
& \ s.t. \ KH=I.
\end{align*}
\thlabel{problem:MVUE}
\end{prob}

Due to the singularity of the covariance matrix~$V$, the closed-form solution is obtained based on the results on Gauss-Markov estimate by~\cite{albert1973gauss}, of which the main result is stated in the following lemma.

\begin{lem}[MVUE with singular covariances]
When the covariance~$V \in \mathbb{R}^{n \times n}$ is a singular matrix, the optimal solution to~\thref{problem:MVUE} is
\begin{equation}
K^* = H^{\dagger} [I - (L V L)^{\dagger} L V]',
\end{equation}
where~$L := I - H H^{\dagger}$.
\thlabel{Lemma:MVUE_optimal}
\end{lem}

The structural similarity between the fusion estimation in \thref{problem:sensor_fusion} and MVUE in~\thref{problem:MVUE} motivates us to find an optimal linear fusion coefficient in closed-form.

\begin{thm}[Closed-form solution to~\thref{problem:sensor_fusion}]
The optimal fusion coefficients in the unbiased linear state fusion in \thref{problem:sensor_fusion} can be expressed as
\begin{equation}
\bm{W}_k^* = [I_{(nN)} - (M \Sigma M)^{\dagger} M \Sigma] \bm{V}_o'^{\dagger},
\end{equation}
where~$M := I_{(nN)} - \bm{V}_o \bm{V}_o^{\dagger}$.
\thlabel{THM:closed_form}
\end{thm}

\begin{pf}
Based on the comparison between~\thref{problem:sensor_fusion} and~\thref{problem:MVUE}, the result follows directly from~\thref{Lemma:MVUE_optimal}.~\qedsymbol
\end{pf}


\section{Simulation}

In this section, we consider a linearized model of inverted pendulum~\cite{messner1999control} to verify the performance of the proposed unbiased linear fusion scheme. The following parameters are chosen:
\begin{itemize}
\item Mass of the cart $M=0.5 \ Kg$;
\item Mass of the pendulum $m=0.2 \ Kg$;
\item Coefficient of friction for cart $b = 0.1 \ N/m/s$;
\item Length to pendulum center of mass $l=0.1 \ m$;
\item Moment of inertia of the pendulum $J= 1 \ Kg \cdot m^2$.
\end{itemize}

We denote the cart's displacement as $x$ and the pendulum angle as $\phi$, i.e., the deviation of the pendulum's position from equilibrium. The following continuous-time dynamic equation is obtained,
\begin{align*}
\begin{bmatrix}
\dot{x}\\ \ddot{x}\\ \dot{\phi}\\ \ddot{\phi}
\end{bmatrix}
=
\begin{bmatrix}
0 & 1 & 0 & 0\\
0 & \frac{-(J + m l^2) b}{p} & \frac{m^2 g l^2}{Jp} & 0\\
0 & 0 & 0 & 1\\
0 & \frac{-mlb}{p} & \frac{mgl(M+m)}{p} & 0
\end{bmatrix}
\begin{bmatrix}
x\\ \dot{x}\\ \phi\\ \dot{\phi}
\end{bmatrix}+
\begin{bmatrix}
0\\
\frac{J + m l^2}{p}\\
0\\
\frac{ml}{p}
\end{bmatrix}
u,
\end{align*}
where $p:=J(M+m) + M m l^2$ and $u$ is the external force.

To discretize the system dynamics, we apply zero-order holding (ZOH) with sampling time $T_s = 0.001 s$ and obtain the discrete-time linear dynamic model as~$X_{k+1} = A X_k + B u_k$, where $X_k := [x_k \ \dot{x}_k \ \phi_k \ \hat{\phi}_k]'$ is the state variable, and $u_k$ is the control input.

In this simulation example, we assume that a random perturbation force is imposed on the cart. Specifically, we model the control input as a Gaussian white noise $u_k \sim \mathcal{N}(0,\sigma^2)$ with $\sigma^2 = 10$. Define~$Q:=\mathds{E}[B u_k u_k' B']=\sigma^2 B B' \geq 0$,then the dynamics of the linear process becomes
\begin{equation}
X_{k+1} = A X_k + w_k,
\end{equation}
where the Gaussian white noise $w_k := B u_k \sim \mathcal{N}(0,Q)$ and $\mathds{E}[w_k w_j']=0$ for any $k \neq j$.

Different sensing technologies are employed to measure the states of the inverted pendulum, e.g., infrared rays and mechanical sensors, etc. Suppose the measurement functions are
\begin{equation}
y_k^{(i)} = C_i X_k + v_k^{(i)}, \ i=1,2,\ldots,10,
\label{eq:sim_observation}
\end{equation}
where $v_k^{(i)} \sim \mathcal{N}(0,R_i)$ with $R_i > 0$ and $\mathds{E}[v_k^{(i)} v_j^{(i)'}]=0$ for $i=1,2,\ldots,10$ and any $k \neq j$.

In~(\ref{eq:sim_observation}), the observation matrices are chosen as
\begin{align*}
&C_1 = \begin{bmatrix}
0 & 0.2 & 0 & 0
\end{bmatrix},\
C_2 = \begin{bmatrix}
1 & 0.5 & 0 & 0\\
0.5 & 1 & 0 & 0
\end{bmatrix},\
C_3=\begin{bmatrix}
1 & 0 & 0 & 0
\end{bmatrix}, \\
&C_4 = \begin{bmatrix}
0 & 0 & 0.5 & 0
\end{bmatrix},\
C_5 = \begin{bmatrix}
1 & 0 & 0 & 0.4\\
0.2 & 0 & 0 & 1
\end{bmatrix},\
C_6=\begin{bmatrix}
0 & 0 & 0 & 1
\end{bmatrix}, \\
&C_7 = \begin{bmatrix}
0 & 0 & 1 & 0
\end{bmatrix},\
C_8 = \begin{bmatrix}
0 & 0 & 1 & 0
\end{bmatrix},\
C_9=\begin{bmatrix}
0 & 1 & 0 & 0.4\\
0 & 0.5 & 0 & 1
\end{bmatrix}, \\
&C_{10} = \begin{bmatrix}
0 & 0 & 0 & 1
\end{bmatrix},
\end{align*}
and the error covariance matrices of the observations noise are
\begin{align*}
& R_1 = 0.04, \ R_2= \textbf{Diag}(0.02, 0.01), \ R_3 = 0.16,\\
& R_4 = 0.01, \ R_5= \textbf{Diag}(0.04, 0.01), \ R_6 = 0.35,\\
& R_7 = 0.02, \ R_8= 0.25, \ R_9 = \textbf{Diag}(0.01, 0.03),\\
& R_{10}=0.09.
\end{align*}

The packet arrival rates of the communication channels for these sensors are:~$[\lambda_1, \ \lambda_2, \ \ldots, \ \lambda_{10}]=$ 
$[0.5, \ 0.6, \ 0.7, \ 0.6, \ 0.7, \ 0.5, \ 0.8, \ 0.5, \ 0.7, \ 0.6]$.

Hence, each sensor is only capable of observing a certain ``sub-component" of the system state, whereas the remote state estimator can fuse the local information linearly to obtain a stable global state estimate.

\begin{figure}[t]
	\begin{center}
		\includegraphics[width=0.5\textwidth]{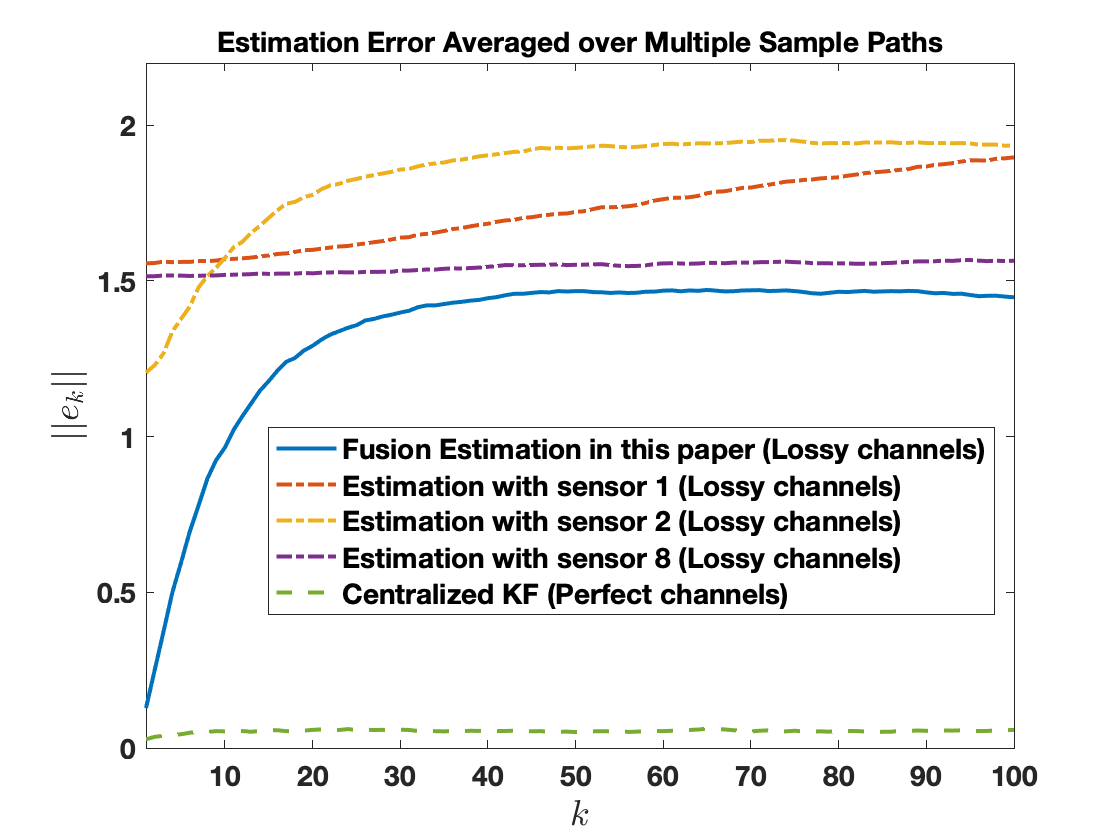}	
		\caption{Numerical simulation of the state estimation error.}
		\label{fig:simulation}
	\end{center}
\end{figure}

It can be verified that \thref{assumption:feasibility} is satisfied, i.e., $(1-\min\limits_{1 \leq i \leq N} \lambda_i) \rho^2(A) = 0.5 \times 1.0004^2 < 1$. The process is simulated for~$50$ sample trajectories, and we obtain the~$2$-norm~$\norm{e_k}$ of the fusion estimation error averaged over all sample paths. The results are shown in Fig.~\ref{fig:simulation}, where the centralized Kalman filter with perfect channels is the benchmark.

\begin{figure}[t]
	\begin{center}
		\includegraphics[width=0.5\textwidth]{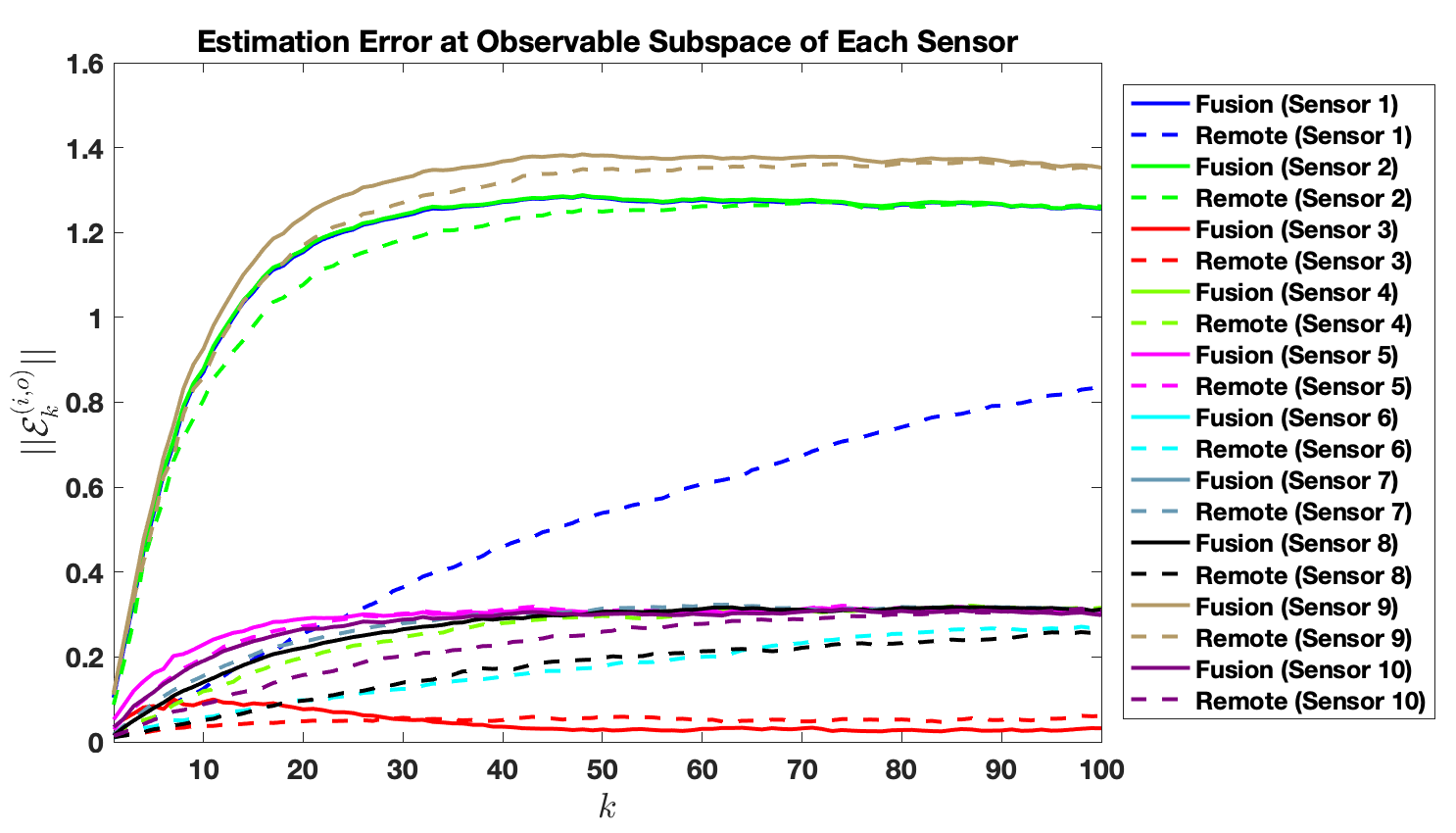}	
		\caption{Error of fusion estimation at the observable subspace of each sensor.}
		\label{fig:subspace}
	\end{center}
\end{figure}

The simulation result indicates that the fusion estimation has achieved a better performance than the estimate generated based only on the observations by some individual sensors; e.g., sensor~$1$, sensor~$2$ or sensor~$8$ as plotted in Fig.~\ref{fig:simulation}, which illustrates the effectiveness of the proposed linear fusion scheme. As mentioned before, the optimal fusion coefficient can be solved in polynomial time~$O(n^5 N^{2.5})$. Therefore, linear fusion can effectively and efficiently integrate information from different sensors. Next, we pay attention to the quality of fusion estimation on the observable subspace of each sensor, as shown in Fig.~\ref{fig:subspace}.


As observed from Fig.~\ref{fig:subspace}, the optimal linear fusion does not necessarily improve the estimation performance in every subspace of the state space. However, since this fusion estimation achieves a better accuracy in the state estimate globally, a tradeoff is achieved among different sensors when calculating the fusion estimation at the remote state estimator. For example, according to the Kalman decomposition~\eqref{eq:Kalman_decomposition}, sensor~$3$ and~$7$ are observing unstable modes of the dynamic system, thus the improvements on the estimation errors in their corresponding observable subspaces are more crucial. Hence, more weights are placed on them while performing linear fusion. On the other hand, sensor~$1$ and~$6$ are observing stable modes of the dynamics, thus information from these sensors may be sacrificed during linear fusion.



\section{Conclusion}
In this paper, we consider an unbiased linear state fusion problem where local observability is not guaranteed for each sensor and the communication channels are lossy. More specifically, each sensor generates an optimal state estimate with a local Kalman filter on its observable subspace, which is then forwarded through lossy channels to the remote state estimator for linear fusion. We propose a networked sensor fusion scheme under collectively observability assumption. The optimal linear fusion coefficients are found through a linear programming. Moreover, the closed-forms expressions of the coefficients are obtained. In the future, globally optimal state fusion scheme as well as sensor fusion in presence of data-injection attacks or eavesdroppers can be considered.


\bibliographystyle{automatica}
\bibliography{reference.bib}

\appendix
\section{Summary of Variable Notations}
\label{appendix:notations}
\begin{table}[H]
	\centering
	\begin{tabularx}{0.5\textwidth}{|l|X|}
	\hline
	Variables & Meanings \\ \hline
	   $x_k$       &       State variable of the linear process         \\ \hline
	   $w_k$       &        Gaussian i.i.d. white noise in the dynamics       \\ \hline
	   $Q$       &          Variance of $w_k$      \\ \hline
	   $y_k^{(i)}$       &    Observation by sensor $i$            \\ \hline
	   $v_k^{(i)}$       &    Gaussian i.i.d. white observation noise at sensor $i$            \\ \hline
	   $R_i$       &        Variance of $v_k^{(i)}$       \\ \hline
	   $\chi_k^{(i,o)}$       &    Projected state variable $x_k$ on the observable subspace of smart sensor~$i$          \\ \hline
	   $\chi_k^{(i)} = [\chi_k^{(i,\overline{o})} \ \chi_k^{(i,o)}]$       &       Equivalent expression of state $x_k$ on a new basis        \\ \hline
	   $A_{i,o}$       &     System matrix on the observable subspace of smart sensor~$i$        \\ \hline
	   $\tilde{w}_k^{(i,o)}$       &        The projection of $w_k$ on the observable subspace of smart sensor~$i$        \\ \hline
	   $\tilde{Q}_{i,o}$       &         Variance of~$\tilde{w}_k^{(i,o)}$     \\ \hline
	   $\hat{\chi}_k^{s,(i)}$       &   The estimate of state $\chi_k^{(i,o)}$ given by the local Kalman filter at sensor~$i$      \\ \hline
	   $\mathcal{E}_k^{s,(i)}$       &        The error of state estimate $\hat{\chi}_k^{s,(i)}$        \\ \hline
	   $\hat{\chi}_{k|k-1}^{s,(i)}$       &  The prediction of $\chi_k^{(i,o)}$ at time $k$             \\ \hline
	   $P_{k|k-1}^{s,(i)}$       &      The error covariance associated with the prediction $\hat{\chi}_{k|k-1}^{s,(i)}$       \\ \hline
	   ${P^*}^{(i,o)}$       &     The steady-state value of $P_{k|k-1}^{s,(i)}$         \\ \hline
	   $P_k^{s,(i)}$       &   The error covariance associated with the state estimate $\hat{\chi}_k^{s,(i)}$             \\ \hline
	   $\overline{P}^{(i,o)}$       &   The steady-state value of $P_k^{s,(i)}$           \\ \hline
	   $\hat{\chi}_k^{(i,o)}$       &      The remote version of the state estimate $\hat{\chi}_k^{s,(i)}$         \\ \hline
	   $\mathcal{E}_k^{(i,o)}$       &       The error of state estimate $\hat{\chi}_k^{(i,o)}$         \\ \hline
	   $P_k^{(ii)}$       &         The error covariance of state estimate $\hat{\chi}_k^{(i,o)}$      \\ \hline
	   $P_k^{(ij)}$       &      The cross-covariance of the error of state estimate $\hat{\chi}_k^{(i,o)}$ and $\hat{\chi}_k^{(j,o)}$           \\ \hline
	   $\hat{x}_k^{(i)}$       &     The projection of $\hat{\chi}_k^{(i,o)}$ back to the original state space with zero-padding on the unobservable modes           \\ \hline
	   $\hat{x}_k$       &     The linear fusion estimation of state variable~$x_k$           \\ \hline
	   $P_k$       &      The error covariance associated with the fusion estimation $\hat{x}_k$          \\ \hline
	   $V_{i,o}$       &     The basis matrix for the observable subspace of sensor~$i$           \\ \hline
	   $W_k^{(i)}$       &        The linear fusion coefficients for weighting the state estimate in the observable subspace of sensor $i$        \\ \hline
	\end{tabularx}

	\end{table}



\section{Proof of~\thref{lemma:Sigma_psd}}
\label{appendix:proof_Sigma_psd}
\begin{pf}
Denote $\bm{\mathcal{E}}_k = [ \mathcal{E}'^{(1,o)}_k \ \mathcal{E}'^{(2,o)}_k \ \ldots \ \mathcal{E}'^{(N,o)}_k ]'$, $\bm{P}_k = [ P_k^{(ij)} ]_{1 \leq i,j \leq N}$ and $V=\bm{\diag} ( V_{1,o} \ V_{2,o} \ \ldots \ V_{N,o} )$, the matrix $\Sigma$ can be expressed as $\Sigma = V \bm{P}_k V'$, where $P_k := \mathbb{E}[\bm{\mathcal{E}}_k \bm{\mathcal{E}}'_k] \geq 0$. By definition, it is known that $P_k^{(ii)}:= \mathbb{E}[\mathcal{E}_k^{(i,o)} \mathcal{E}_k^{(i,o)'}]$ is symmetric for any~$k \in \mathbb{N}_+$ and~$i$. Then, it can be directly obtained that each diagonal block $V_{i,o} P_k^{(ii)} V'_{i,o} \in \mathbb{R}^{n \times n}$ in the matrix $\Sigma$ is symmetric. Hence, it remains to show that $V_{i,o} P_k^{(ij)} V'_{j,o} = (V_{j,o} P_k^{(ji)} V'_{i,o})'$ holds for any off-diagonal elements~$i \neq j$, i.e.,~$P'^{(ji)}_k  = P^{(ij)}_k$. According to~\thref{assumption:ctrl_obser} and~\eqref{eq:cross_correlation}, given that~$Q > 0$ is symmetric, it suffices to show~$\overline{\Gamma}'_{ji} = \overline{\Gamma}_{ij}$.

According to~\thref{lemma:fixed_point_cross_covariance}, it can be concluded that~$\overline{\Gamma}_{ij}$ and~$\overline{\Gamma}_{ji}$ are limits of the sequences~$\{\Gamma_k^{ij}\}_{k \geq 0}$ and~$\{\Gamma_k^{ji}\}_{k \geq 0}$ generated by recursively adopting the operators~$\mathcal{T}_{ij}$ and~$\mathcal{T}_{ji}$ separately. For arbitrary initial values $\Gamma_0^{ij} \in \mathbb{R}^{n_{i,o} \times n_{j,o}}$ and $\Gamma_0^{ji} \in \mathbb{R}^{n_{j,o} \times n_{i,o}}$, with similar arguments as in the proof of~\thref{lemma:fixed_point_cross_covariance}, we can obtain that
\begin{align*}
& \ \ \ \ \norm{\Gamma^{ij}_k - \Gamma'^{ji}_k}
\leq \norm{(A_{i,o} - K_i^* \tilde{C}_i A_{i,o})^k} \cdot \norm{\Gamma^{ij}_0 - \Gamma'^{ji}_0} \cdot\\
& \ \ \ \ \norm{(A_{j,o} - K_j^* \tilde{C}_j A_{j,o})'^k}\\
& \ \ \ \leq [\rho(A_{i,o} - K_i^* \tilde{C}_i A_{i,o}) + \epsilon]^k \cdot \norm{\Gamma^{ij}_0 - \Gamma'^{ji}_0} \cdot\\
& \ \ \ \ [\rho(A_{j,o} - K_j^* \tilde{C}_j A_{j,o}) + \epsilon]^k, \ \forall k > K_0,
\end{align*}
where the last inequality is based on \eqref{eq:spectrum_radius}, and~$
0<\rho(A_{i,o} - K_i^* \tilde{C}_i A_{i,o}) + \epsilon<1$ and~$
0<\rho(A_{j,o} - K_j^* \tilde{C}_j A_{j,o}) + \epsilon<1$.

Hence, for any initial values~$\Gamma^{ij}_0$ and~$\Gamma^{ji}_0$, it can be concluded that~$0 \leq \norm{\overline{\Gamma}_{ij} - \overline{\Gamma}'_{ji}} = \lim\limits_{k \to \infty} \norm{\Gamma^{ij}_k - \Gamma'^{ji}_k}=0$, i.e.,~$\overline{\Gamma}_{ij} = \overline{\Gamma}'_{ji}$.

Therefore, the symmetric matrix $\Sigma = V \bm{P}_k V' \geq 0$. \qedsymbol
\end{pf}

\section{Proof of~\thref{lemma:equivalence_fusion_SDP}}
\label{appendix:proof_equivalence_fusion_SDP}
\begin{pf}
As~$\bm{X}_k \succeq \bm{W}_k \bm{W}_k' \succeq 0$, there is~$\bm{X}_k - \bm{W}_k \bm{W}_k' \succeq 0$. For any pair of optimal solution~$\left(\bm{W}_k^*,\bm{X}_k^*\right)$, we can construct an auxiliary variable pair~$\left( \bm{W}_k^*,\tilde{\bm{X}}_k^* \right)$ such that $\tilde{\bm{X}}_k^* = \bm{W}_k^* \bm{W}_k'^*$. Hence, the constraints are still satisfied. Moreover, the value of the objective function in \thref{problem:SDP} achieved under the auxiliary variable pair is
\begin{align*}
& \tr ( \Sigma \tilde{\bm{X}}_k^* ) = \tr \left( \Sigma (\tilde{\bm{X}}_k^* - \bm{W}_k^* \bm{W}_k'^*) \right) + \tr ( \Sigma \bm{W}_k^* \bm{W}_k'^* )\\
&= 0 + \tr (\Sigma^{\frac{1}{2}} \bm{W}_k^* \bm{W}_k'^* \Sigma^{\frac{1}{2}} ) \leq \tr (\Sigma^{\frac{1}{2}} \bm{X}_k^* \Sigma^{\frac{1}{2}} ) = \tr ( \Sigma \bm{X}_k^* ),
\end{align*}
where the inequality is due to the constraint~$\bm{X}_k \succeq \bm{W}_k \bm{W}_k'$.

Since the pair~$\left( \bm{W}_k^*, \bm{X}_k^* \right)$ is an optimal solution, there should be~$\tr \left( \Sigma \bm{X}_k^* \right) \leq \tr \left( \Sigma \tilde{\bm{X}}_k^* \right)$. Hence, we obtain that~$\tr \left( \Sigma \tilde{\bm{X}}_k^* \right)=\tr \left( \Sigma \bm{X}_k^* \right)$, i.e., the optimal value of the objective function in \thref{problem:SDP} can be achieved with strict equality~$\bm{X}_k = \bm{W}_k \bm{W}_k'$.

Thus, there is no loss of optimality restricting the feasible decision variable in \thref{problem:SDP} to be constrained by the equality~$\bm{X}_k = \bm{W}_k \bm{W}_k'$. This gives~$\tr \left( \Sigma \bm{X}_k \right) = \tr \left( \Sigma \bm{W}_k \bm{W}_k' \right) = \tr \left( \bm{W}_k' \Sigma \bm{W}_k \right) = \tr \left( P_k \right)$. In other words, the optimal solution~$\bm{W}_k^*$ to~\thref{problem:sensor_fusion} will coincide with the optimal solution to \thref{problem:SDP}. \qedsymbol
\end{pf}

\section{Proof of~\thref{THM:optimal_coeff}}
\label{appendix:proof_optimal_coeff}
\begin{pf}
In order to transform the original~\thref{problem:sensor_fusion} equivalently as a linear programming~(LP), we find the dual problem of \thref{problem:SDP} and verify that strong duality holds.

Denote the Lagrangian multiplier associated with $\bm{W}_k' \bm{V}_o = I_{(n)}$ as~$\Lambda_1 \in \mathbb{R}^{n \times n}$. The one associated with~$\bm{X}_k \succeq \bm{W}_k \bm{W}_k'$ is denoted as~$\Lambda_2 \in \mathbb{R}^{nN \times nN}$, which satisfies~$\Lambda_2 \succeq 0$. Hence, the Lagrangian is expressed as
\begin{align*}
L \left( \bm{W}_k, \bm{X}_k, \Lambda_1, \Lambda_2 \right) &= \tr\left( \Sigma \bm{X}_k \right) - \tr\left(\Lambda_1' (\bm{W}_k' \bm{V}_o - I_{(n)}) \right)\\
& \ \ \ \ -\tr\left( \Lambda_2' (\bm{X}_k - \bm{W}_k \bm{W}_k') \right).
\end{align*}

Now, we apply Karush-Kuhn-Tucker (KKT) condition to characterize the optimal solution to \thref{problem:SDP}.
\begin{align*}
\begin{cases}
\frac{\partial}{\partial \bm{W}_k} L \left( \bm{W}_k, \bm{X}_k, \Lambda_1, \Lambda_2 \right) = 0; \cr
\frac{\partial}{\partial \bm{X}_k} L \left( \bm{W}_k, \bm{X}_k, \Lambda_1, \Lambda_2 \right) = 0; \cr
\bm{W}_k' \bm{V}_o = I_{(n)}; \cr
\bm{X}_k \succeq \bm{W}_k \bm{W}_k'; \cr
\Lambda_2 \succeq 0; \cr
\tr\left( \Lambda_2' ( \bm{X}_k - \bm{W}_k \bm{W}_k') \right) = 0.
\end{cases}
\end{align*}

Through manipulating the KKT conditions, we obtain that
\begin{align*}
\Lambda_2 = \Sigma; \
2 \Sigma \bm{W}_k = \bm{V}_o \Lambda_1'; \
\bm{W}_k' \bm{V}_o = I_{(n)}.
\end{align*}

Hence, the objective of the dual problem of \thref{problem:SDP} will be
\begin{align*}
& \ \ \ \ \max\limits_{\Lambda_1, \Lambda_2} \min\limits_{\bm{W}_k, \bm{X}_k} L \left( \bm{W}_k, \bm{X}_k, \Lambda_1, \Lambda_2 \right)\\
&= \max\limits_{\Lambda_1, \Lambda_2} \tr \left( \Lambda_1 \right) + \tr \left( \Lambda_2' \bm{W}_k \bm{W}_k' \right) - \tr \left( \Lambda_1' \bm{W}_k' \bm{V}_o \right)\\
&= \max\limits_{\Lambda_1, \Lambda_2} \frac{1}{2} \tr \left( \bm{V}_o \Lambda_1' \bm{W}_k' \right) = \max\limits_{\Lambda_1} \frac{1}{2} \tr \left( \Lambda_1 \right).
\end{align*}

Thus, the dual of \thref{problem:SDP} is~\thref{problem:dual}. It remains to check the strong duality.

From \thref{problem:SDP}, the objective~$\tr \left( \Sigma \bm{X}_k \right)$ and the constraint~$\bm{W}_k' \bm{V}_o = I_{(n)}$ are linear, while the matrix inequality constraint~$\bm{X}_k \succeq \bm{W}_k \bm{W}_k'$ can be equivalently expressed as
\begin{align*}
\begin{bmatrix}
I_{(n)} & \bm{W}_k'\\
\bm{W}_k & \bm{X}_k
\end{bmatrix} \succeq 0,
\end{align*}
which is a convex constraint.

For a variable~$\bm{W}_k$ satisfying~$\bm{W}_k' \bm{V}_o = I_{(n)}$, if we pick~$\bm{X}_k = \bm{W}_k \bm{W}_k' + \epsilon I_{(nN)}$ with~$\epsilon > 0$, the Slater's condition hold, i.e.,~$\bm{W}_k' \bm{V}_o = I_{(n)}$ and~$\bm{X}_k \succ \bm{W}_k \bm{W}_k'$ are satisfied at the same time. Therefore, strong duality is verified, hence \thref{problem:sensor_fusion} is equivalent to \thref{problem:dual}. \qedsymbol
\end{pf}

\section{Proof of~\thref{prop:stable}}
\label{appendix:proof_stable}
\begin{pf}
As shown in~\thref{THM:optimal_coeff}, the optimal solution to~\thref{problem:sensor_fusion} is equivalent to \thref{problem:SDP}. Denote its optimal solution as~$\left( \bm{W}_k^*, \bm{X}_k^* \right)$, then for any pair of feasible variables~$\left( \bm{W}_k, \bm{X}_k \right)$, it can be obtained that for any~$\bm{X}_k$ the error covariance~$P_k^*$ under optimal fusion coefficients satisfies
\begin{equation}
\tr \mathbb{E}[P_k^*] = \mathbb{E} [\tr \left( P_k^* \right)] = \mathbb{E}[\tr \left( \Sigma \bm{X}_k^* \right)] \leq \mathbb{E}[\tr \left( \Sigma \bm{X}_k \right)].
\label{eq:upper_bound_Pk}
\end{equation}

Hence, it suffices to show that~$\mathbb{E}[\tr\left( \Sigma \bm{X}_k \right)] < \infty$.

Based on~\thref{assumption:ctrl_obser}, the linear process is collectively observable by the~$N$ sensors. Hence, we have $\rank \left[ V_{1,o} \ V_{2,o} \ \ldots \ V_{N,o} \right] = n$, as shown below by contradiction.

Assume that $\rank [ V_{1,o} \ V_{2,o} \ \ldots \ V_{N,o}] < n$, then there exists a non-zero vector~$\bm{v} \in \mathbb{R}^{\sum_{i=1}^N n_{i,o}} / \{\bm{0}\}$ such that~$\left[ V_{1,o} \ V_{2,o} \ \ldots \ V_{N,o} \right] \bm{v} = \bm{0} \in \mathbb{R}^n$. As~$\bm{v} \neq \bm{0}$, there exists a sensor~$i_0$ with $V_{i,o} \bm{v}_i = \bm{0}$ for a non-zero~$\bm{v}_i \in \mathbb{R}^{n_{i,o}} / \{\bm{0}\}$, hence~$\rank V_{i,o} < n_{i,o}$, which contradicts with~$V_{i,o}' V_{i,o} = I_{(n_{i,o})}$ given by Kalman decomposition. Consequently, we have $\rank [ V_{1,o} \ V_{2,o} \ \ldots \ V_{N,o} ] = n$.

The rank of matrix~$\bm{V}_o \in \mathbb{R}^{nN \times n}$ satisfies
\begin{align*}
n \geq \rank \bm{V}_o & \geq \rank \left( \bm{\diag} (V_{1,o}', V_{2,o}', \ldots, V_{N,o}') \cdot \bm{V}_o \right)\\
&=\rank \left[ V_{1,o} \ V_{2,o} \ \ldots \ V_{N,o} \right]' = n,
\end{align*}
i.e., the matrix~$\bm{V}_o$ is of full row rank. Thus, the constraint~$\bm{W}_k \bm{V}_o = I_{(n)}$ is feasible.

Now, we arbitrarily pick and fix a $\overline{\bm{W}}$ satisfying~$\overline{\bm{W}} \bm{V}_o = I_{(n)}$. Based on Schur decomposition, we can obtain that~$\overline{\bm{W}} \ \overline{\bm{W}}' \preceq \overline{m} I_{(nN)}$ where~$\overline{m} > \sigma^2_{\max} (\overline{\bm{W}})$, i.e, the square of the maximum singular value of~$\overline{\bm{W}}$. We choose~$\overline{\bm{X}} = \overline{m} \cdot I_{(nN)}$, then the pair~$\left( \overline{\bm{W}}, \overline{\bm{X}} \right)$ is feasible in~\eqref{eq:obj_SDP} at any time~$k$.

According to~\eqref{eq:upper_bound_Pk}, it can be concluded that
\begin{align*}
& \ \ \ \tr \mathbb{E}[P_k^*] \leq \mathbb{E}[\tr \left( \Sigma \overline{\bm{X}} \right)] =\overline{m} \cdot \mathbb{E}[\tr \left( \Sigma \right)]\\
&= \overline{m} \cdot \mathbb{E}[\sum\limits_{i=1}^N \tr \left( V_{i,o}' V_{i,o} P_k^{(ii)} \right)] = \overline{m} \cdot \mathbb{E}[\sum\limits_{i=1}^N \tr \left( P_k^{(ii)} \right)]\\
&\leq \overline{m} \cdot \sum\limits_{i=1}^N n_{i,o} \cdot \bigg\{ \mathbb{E} [\rho^{2 \tau_k^{(i)}} \left( A_{i,o} \right)] \cdot\\
& \ \ \  \bigg[\tr\left(\overline{P}^{(i,o)}\right) - \frac{\tr\left(V'_{i,o} Q V_{i,o}\right)}{1-\rho^2 \left( A_{i,o} \right)} \bigg]+ \frac{\tr\left(V'_{i,o} Q V_{i,o}\right)}{1-\rho^2 \left( A_{i,o} \right)} \bigg\}\\
&< \infty.
\end{align*}

The last inequality holds according to~\thref{assumption:feasibility}, as
\begin{align*}
\mathbb{E} [\rho^{2 \tau_k^{(i)}} \left( A_{i,o} \right)] &= \sum\limits_{k=0}^{\infty} \rho^{2k} \left( A_{i,o} \right) \cdot \mathbb{P} \left( \tau_k^{(i)} = k \right)\\
&= \sum\limits_{k=0}^{\infty} \rho^{2k} \left( A_{i,o} \right) \cdot \lambda_i (1 - \lambda_i)^k\\
&\leq \lambda_i \sum\limits_{k=0}^{\infty} \bigg[\left(1 - \min\limits_{1 \leq i \leq N} \lambda_i \right) \rho^2 (A) \bigg]^k\\
&<\infty.
\end{align*}

Therefore, the remote state estimate is stable. \qedsymbol
\end{pf}

\par\noindent 
\parbox[t]{\linewidth}{
\noindent
\noindent {\bf Yuchi Wu}\
received the B.Eng. degree from School of Electronic and Information Engineering, Beihang University, Beijing, China, in 2015. Later, he received the M.Phil. degree in 2017 and the Ph.D. degree in 2020, both from Department of Electronic and Computer Engineering, Hong Kong University of Science and Technology, Kowloon, Hong Kong. From March 2020 to May 2020, he was a visiting student in the School of Mathematics and Statistics, Carleton University, Ottawa, ON, Canada. His research interests include mean-field games, large-scale systems and state estimation in wireless sensor networks.}
\vspace{4\baselineskip}

\par\noindent 
\parbox[t]{\linewidth}{
\noindent
\noindent {\bf Kemi Ding}\
received the B.S. degree in Electronic and Information Engineering from Huazhong University of Science and Technology, China, in 2014 and the Ph.D. degree in the Department of Electronic and Computer Engineering from Hong Kong University of Science and Technology in 2018. She is currently a postdoctoral researcher at the School of Electrical and Electronic Engineering, Nanyang Technological University, Singapore. Prior to this, she was a postdoctoral researcher in the School of Electrical, Computer and Energy Engineering, Arizona State University from September 2018 to August 2019. Her current research interests include graph signal processing, cyber-physical system security/privacy, networked state estimation, and game theory.
}
\vspace{4\baselineskip}

\par\noindent 
\parbox[t]{\linewidth}{
\noindent
\noindent {\bf Yuzhe Li}\
is currently a Professor in the State Key Laboratory of Synthetical Automation for Process Industries, Northeastern University, Shenyang, China. He received the B.S. degree in Mechanics from Peking University, China in 2011 and the Ph.D. degree in Electronic and Computer Engineering from the Hong Kong University of Science and Technology (HKUST) in 2015. Between June 2013 and August 2013, he was a visiting scholar in the University of Newcastle, Australia. From September 2015 to September 2017, he was a Postdoctoral Fellow at the Department of Electrical and Computer Engineering, University of Alberta, Canada. His research interests include cyber-physical systems security, sensor power control and networked state estimation.
}
\vspace{4\baselineskip}

\par\noindent 
\parbox[t]{\linewidth}{
\noindent
\noindent {\bf Ling Shi}\
received the B.E. degree in electrical and electronic engineering from Hong Kong University of Science and Technology, Kowloon, Hong Kong, in 2002 and the Ph.D. degree in Control and Dynamical Systems from California Institute of Technology, Pasadena, CA, USA, in 2008. He is currently a Professor in the Department of Electronic and Computer Engineering, and the associate director of the Robotics Institute, both at the Hong Kong University of Science and Technology. His research interests include cyber-physical systems security, networked control systems, sensor scheduling, event-based state estimation and exoskeleton robots. He is a senior member of IEEE. He served as an editorial board member for the European Control Conference 2013-2016. He was a subject editor for International Journal of Robust and Nonlinear Control (2015-2017), an associate editor for IEEE Transactions on Control of Network Systems (2016-2020), an associate editor for IEEE Control Systems Letters (2017-2020), and an associate editor for a special issue on Secure Control of Cyber Physical Systems in IEEE Transactions on Control of Network Systems (2015-2017). He also served as the General Chair of the 23rd International Symposium on Mathematical Theory of Networks and Systems (MTNS 2018). He is a member of the Young Scientists Class 2020 of the World Economic Forum (WEF).}
\vspace{4\baselineskip}

\end{document}